



\font\twelverm=cmr10 scaled 1200    \font\twelvei=cmmi10 scaled 1200
\font\twelvesy=cmsy10 scaled 1200   \font\twelveex=cmex10 scaled 1200
\font\twelvebf=cmbx10 scaled 1200   \font\twelvesl=cmsl10 scaled 1200
\font\twelvett=cmtt10 scaled 1200   \font\twelveit=cmti10 scaled 1200

\skewchar\twelvei='177   \skewchar\twelvesy='60


\def\twelvepoint{\normalbaselineskip=12.4pt
  \abovedisplayskip 12.4pt plus 3pt minus 9pt
  \belowdisplayskip 12.4pt plus 3pt minus 9pt
  \abovedisplayshortskip 0pt plus 3pt
  \belowdisplayshortskip 7.2pt plus 3pt minus 4pt
  \smallskipamount=3.6pt plus1.2pt minus1.2pt
  \medskipamount=7.2pt plus2.4pt minus2.4pt
  \bigskipamount=14.4pt plus4.8pt minus4.8pt
  \def\rm{\fam0\twelverm}          \def\it{\fam\itfam\twelveit}%
  \def\sl{\fam\slfam\twelvesl}     \def\bf{\fam\bffam\twelvebf}%
  \def\mit{\fam 1}                 \def\cal{\fam 2}%
  \def\tt{\twelvett}
  \textfont0=\twelverm   \scriptfont0=\tenrm   \scriptscriptfont0=\sevenrm
  \textfont1=\twelvei    \scriptfont1=\teni    \scriptscriptfont1=\seveni
  \textfont2=\twelvesy   \scriptfont2=\tensy   \scriptscriptfont2=\sevensy
  \textfont3=\twelveex   \scriptfont3=\twelveex  \scriptscriptfont3=\twelveex
  \textfont\itfam=\twelveit
  \textfont\slfam=\twelvesl
  \textfont\bffam=\twelvebf \scriptfont\bffam=\tenbf
  \scriptscriptfont\bffam=\sevenbf
  \normalbaselines\rm}




\def\beginlinemode{\endmode
  \begingroup\parskip=0pt \obeylines\def\\{\par}\def\endmode{\par\endgroup}}
\def\beginparmode{\endmode
  \begingroup \def\endmode{\par\endgroup}}
\let\endmode=\par
{\obeylines\gdef\
{}}
\def\singlespace{\baselineskip=\normalbaselineskip}

\def\oneandahalfspace{\baselineskip=\normalbaselineskip
  \multiply\baselineskip by 3 \divide\baselineskip by 2}
\def\doublespace{\baselineskip=\normalbaselineskip \multiply\baselineskip by 2}

\newcount\firstpageno
\firstpageno=2
\footline={\ifnum\pageno<\firstpageno{\hfil}\else{\hfil\twelverm\folio\hfil}\fi}
\let\rawfootnote=\footnote		
\def\footnote#1#2{{\rm\singlespace\parindent=0pt\rawfootnote{$^#1$}{#2}}}
\def\raggedcenter{\leftskip=4em plus 12em \rightskip=\leftskip
  \parindent=0pt \parfillskip=0pt \spaceskip=.3333em \xspaceskip=.5em
  \pretolerance=9999 \tolerance=9999
  \hyphenpenalty=9999 \exhyphenpenalty=9999 }
\def\dateline{\rightline{\ifcase\month\or
  January\or February\or March\or April\or May\or June\or
  July\or August\or September\or October\or November\or December\fi
  \space\number\year}}
\def\received{\vskip 3pt plus 0.2fill
 \centerline{\sl (Received\space\ifcase\month\or
  January\or February\or March\or April\or May\or June\or
  July\or August\or September\or October\or November\or December\fi
  \qquad, \number\year)}}

\hsize=6.5truein
\hoffset=0truein
\vsize=8.9truein
\voffset=0truein
\parskip=\medskipamount
\twelvepoint		
\doublespace		
\overfullrule=0pt	



\def\title			
  {\null\vskip 3pt plus 0.2fill
   \beginlinemode \singlespace \raggedcenter \bf}

\def\author			
  {\vskip 3pt plus 0.2fill \beginlinemode
   \singlespace \raggedcenter}

\def\affil			
  {\vskip 3pt plus 0.1fill \beginlinemode
   \singlespace \raggedcenter \sl}

\def\abstract			
  {\vskip 3pt plus 0.3fill \beginparmode
   \singlespace \narrower ABSTRACT: }

\def\endtitlepage		
  {\endpage			
   \body}

\def\body			
  {\beginparmode}		

\def\head#1{			
  \goodbreak\vskip 0.5truein	
  {\immediate\write16{#1}
   \raggedcenter \uppercase{#1}\par}
   \nobreak\vskip 0.25truein\nobreak}

\def\refto#1{$^{#1}$}		

\def\references			
  {\head{References}		
   \beginparmode
   \frenchspacing \parindent=0pt \leftskip=1truecm
   \parskip=8pt plus 3pt \everypar{\hangindent=\parindent}}

\gdef\refis#1{\indent\hbox to 0pt{\hss#1.~}}	

\gdef\journal#1, #2, #3, 1#4#5#6{		
    {\sl #1~}{\bf #2}, #3 (1#4#5#6)}		

\def\refstylenp{		
  \gdef\refto##1{ [##1]}				
  \gdef\refis##1{\indent\hbox to 0pt{\hss##1)~}}	
  \gdef\journal##1, ##2, ##3, ##4 {			
     {\sl ##1~}{\bf ##2~}(##3) ##4 }}

\def\refstyleprnp{		
  \gdef\refto##1{ [##1]}				
  \gdef\refis##1{\indent\hbox to 0pt{\hss##1)~}}	
  \gdef\journal##1, ##2, ##3, 1##4##5##6{		
    {\sl ##1~}{\bf ##2~}(1##4##5##6) ##3}}

\def\pl{\journal Phys.\ Lett., }

\def\prd{\journal Phys.\ Rev.\ D, }

\def\prl{\journal Phys.\ Rev.\ Lett., }

\def\rmp{\journal Rev.\ Mod.\ Phys., }

\def\figurecaptions		
  {\endpage
   \beginparmode
   \head{Figure Captions}
}

\def\endpage			
  {\vfill\eject}

\def\endpaper			
  {\endmode\vfill\supereject}

%
\def\Rome#1{\uppercase\expandafter{\romannumeral#1}}
\def\rome#1{\lowercase\expandafter{\romannumeral#1}}

\def\1{$^1$}
\def\2{$^2$}
\def\3{$^3$}
\def\4{$^4$}
\def\5{$^5$}
\def\6{$^6$}
\def\7{$^7$}
\def\8{$^8$}
\def\9{$^9$}
\def\0{$^0$}

\newcount\refno  
\refno=0 \def\nextref{\advance\refno by 1 $^{\the\refno)}$}
\def\nextrefno{\advance\refno by 1 $^{\the\refno}$}  
\newcount\tempref
\def\refs#1{\tempref=\refno \advance\refno by #1 \advance\tempref by 1
    {$^{\the\tempref-\the\refno)}$}}
\def\tworefs{\tempref=\refno \advance\refno by 2 \advance\tempref by 1
    {$^{\the\tempref,\the\refno)}$}}
\def\refsno#1{\tempref=\refno \advance\refno by #1 \advance\tempref by 1
    {$^{\the\tempref-\the\refno}$}}  
\def\tworefsno{\tempref=\refno \advance\refno by 2 \advance\tempref by 1
    {$^{\the\tempref,\the\refno}$}}  
\newcount\figno  
\figno=0 \def\nextfig{\advance\figno by 1 {Fig.\ \the\figno}}
\newcount\tempfig
\def\figs#1{\tempfig=\figno \advance\figno by #1 \advance\tempfig by 1
    {Figs.\ \the\tempfig--\the\figno}}
\def\twofigs{\tempfig=\figno \advance\figno by 2 \advance\tempfig by 1
    {Figs.\ \the\tempfig\ and\ \the\figno}}
\def\pagefigs#1{\vskip.3cm\singlespace\noindent{\global\advance\figno by 1
     {Fig.\ \the\figno}. \ #1}}
\def\today{\number\day\space\ifcase\month\or January\or February\or March\or
April\or May\or June\or July\or August\or September\or October\or November\or
December\fi\space\number\year}
\def\overrharpup#1{\vbox{\ialign{##\crcr
    $\rightharpoonup$\crcr\noalign{\kern-1pt\nointerlineskip}
    $\hfil\displaystyle{#1}\hfil$\crcr}}}
\def\overrharpdown#1{\vbox{\ialign{##\crcr
    $\rightharpoondown$\crcr\noalign{\kern-1pt\nointerlineskip}
    $\hfil\displaystyle{#1}\hfil$\crcr}}}

\def\ref#1{Ref. #1}			
\def\frac#1#2{{\textstyle{#1 \over #2}}}

\def\sla{\raise.15ex\hbox{$/$}\kern-.57em}
\def\leaderfill{\leaders\hbox to 1em{\hss.\hss}\hfill}
\def\twiddle{\lower.9ex\rlap{$\kern-.1em\scriptstyle\sim$}}
\def\bigtwiddle{\lower1.ex\rlap{$\sim$}}
\def\gtwid{\mathrel{\raise.3ex\hbox{$>$\kern-.75em\lower1ex\hbox{$\sim$}}}}
\def\ltwid{\mathrel{\raise.3ex\hbox{$<$\kern-.75em\lower1ex\hbox{$\sim$}}}}
\def\square{\kern1pt\vbox{\hrule height 1.2pt\hbox{\vrule width 1.2pt\hskip 3pt
   \vbox{\vskip 6pt}\hskip 3pt\vrule width 0.6pt}\hrule height 0.6pt}\kern1pt}

\singlespace
\def\nue{$\nu_e$}
\def\num{$\nu_\mu$}
\def\nut{$\nu_\tau$}

\rightline{UCSB--HEP--95--1}
\rightline{UMD--PP--95--107}
\rightline{March, 1995}
\vskip 5mm
\centerline{\bf  An Inverted Neutrino Mass Hierarchy for Hot Dark Matter}
\centerline{\bf  and the Solar Neutrino Deficit}
\vskip 1cm
\centerline{{\bf David O. Caldwell}\footnote*{Work supported by the Department
of Energy grant No. DE-EG03-91ER40618}}
\centerline{Department of Physics, University of California, Santa Barabara,
California 93106}
\vskip 5mm
\centerline{{\bf Rabindra N. Mohapatra}\footnote\dagger
{Work supported by the National
Science Foundation grant No.PHY9421385}}
\centerline{Department of Physics, University of Maryland, College Park,
Maryland 20742}
\vskip 1cm
\centerline{ABSTRACT}

{\singlespace\narrower\narrower
The cosmological model in which $20\%$ of the dark matter is
shared by two nearly equal mass neutrinos fits the structure of the
universe on all scales. This has been motivated by a
$\nu_{\mu}\rightarrow \nu_{\tau}$
oscillation explanation of the deficit of atmospheric $\nu_{\mu}$'s.
If the observed atmospheric ${{\nu_{\mu}}\over{\nu_e}}$ ratio
has an alternative explanation, the cosmological model can be
retained if the deficit of solar neutrinos is explained by
$\nu_e \rightarrow\nu_{\tau}$ oscillation. In this case, an inverted neutrino
mass hierarchy is required with $m_{\nu_\mu}\ll m_{\nu_e}\approx m_{\nu_\tau}
\approx 2.4$ eV.
We show that if there exists an $L_e-L_{\tau}$ symmetry in nature,
then both the near mass degeneracy of $\nu_e$ and $\nu_{\tau}$,
as well as the consistency of the above mass values for neutrino masses
 with the negative results of the neutrinoless double beta
decay search experiments are easily understood. We show that
this symmetry implemented in the context of
a high-scale left-right symmetric model with the see-saw
mechanism can lead to a simple
theoretical understanding of the desired form of the mass matrix.

\vfill\eject
\oneandahalfspace

\noindent{\bf 1. Introduction}

The standard electroweak model of Glashow, Weinberg and Salam predicts
that the neutrinos are massless. There are, however, strong indications
from solar neutrino data that the only way to reconcile all four
experimental results[1] with the calculations
of the standard solar model[2]
is to assume that the neutrinos have mass and they mix among themselves.
There are also indications of a nonzero neutrino mass from
attempts to understand the observed large scale structure in the universe.
The best fit of any available model to the structure on all
scales, such as the anisotropy of cosmic microwave background,
galaxy-galaxy angular correlations, velocity fields on large and small
scales, correlations of galaxy clusters, etc. seems
to be provided by the assumption that
the dark matter in the universe is made up of about
$75\%$ cold (CDM) and about $20\%$ hot dark matter (HDM).
The obvious HDM candidate is a neutrino with mass
in the few eV range. It has recently been argued that actually
two neutrino species degenerate in mass ($\simeq 2.4$ eV) provide a
better detailed fit to data than a single species with the same total
mass[3].

 Another observation in support of a nonzero neutrino mass
comes from attempts to understand a possible deficit of muon neutrinos
in the data on cosmic ray neutrinos observed in several recent
experiments[4]. The conclusion of a neutrino mass from these latter
experiments may be on a somewhat weaker footing, since
all experiments are not in agreement, and also there exist arguments
that the deficit may have an alternative explanation ( see, for example,
Ref.5). The atmospheric neutrino deficit, explained by $\nu_{\mu}\rightarrow
\nu_{\tau}$ oscillations with two nearly
degenerate $\nu_{\mu}$ and $\nu_{\tau}$
was the original motivation for the two-neutrino cold plus hot dark matter
model[3]. Because of the success of that model, we pursue here
an alternative neutrino mass scenario which could preserve the model
even if the observed ${{\nu_\mu}\over{\nu_e}}$ ratio does not require a
neutrino mass explanation. It will turn out that the neutrino mass
hierarchy needed may have some advantages.

\noindent{\bf 2. Input information}

Of the variety of constraints on possible neutrino masses
and mixings that come from accelerator searches for neutrino oscillations
as well as cosmological and astrophysical considerations,
we briefly summarize those that are directly relevant for our considerations.

\noindent{\it Solar neutrino deficit}:

\noindent Because two out of the three types of
 solar neutrino experiments have to
be wrong for an astrophysical explanation of the deficit to work[6],
it is likely that the explanation of these data involves the oscillation of
the $\nu_e$ to another species of neutrino. The required mass-squared
difference is $\Delta m^2_{ei}\approx 10^{-5}~eV^2$ where $i=\mu~or~\tau$.
The mixing angles favored by data[7] are $sin^22\theta\simeq 0.007$,
the so-called small-angle Mikheyev-Smirnov-Wolfenstein(MSW)
solution and $sin^22\theta\simeq 0.8-1.0$, the
MSW large-angle solution. There is also a  vacuum large-angle solution,
which requires $\Delta m^2_{ei}\simeq 10^{-10}~eV^2$.

\noindent{\it Neutrinoless double beta decay}:

\noindent The latest results from the search for neutrinoless double beta decay
in $^{76}Ge$ seem to imply (with a possibly optimistic set
 of nuclear matrix element
evaluations) that $\langle m_{\nu_e}\rangle\leq .68$ eV at the
$68\%$ confidence
level[8]. Other matrix elements could increase that limit by up to a
factor of two. The effective neutrino mass measured in that process is
$$\langle m_\nu\rangle\approx \Sigma_i\eta_iU_{ei}m_i, \eqno(1)$$
where each neutrino of mass $m_i$ contributes to the total through the mixing
matrix element $U_{ei}$, but with a sign $\eta_i=\pm 1$
determined by its CP-eigen-value, so that cancellations can occur.

\noindent{\it Supernova r-process}:

\noindent It has been noted by Qian
et al[9] that for mass-squared difference $|m^2_{\nu_e}-m^2_{\nu_X}|$
(where $X= \mu~ or~ \tau$) in the $\geq eV^2$ range, an MSW resonance condition
can be met in the supernova environment causing rapid transition of $\nu_\mu$
and $\nu_{\tau}$ to $\nu_e$, provided the mixing angles satisfy the
constraint $sin^2 2\theta\geq 10^{-5}$ or so. Since the $\nu_\mu$ and
$\nu_\tau$ energies are about a factor of two larger than the $\nu_e$ energies,
such transitions generate extra energetic $\nu_e$'s which have a
larger cross-section for $\nu_e+n\rightarrow p+e^-$,
depleting neutrons. The rapid capture  of neutrons (r-process) in the
neutrino-heated ejecta of supernovae is believed to be responsible
for much of the production of the heavy elements; so one infers that for mass
differences in the above range, the mixing angles must be severely restricted.
This constraint, if taken seriously, is therefore very relevant to the
discussion of consistent neutrino mass matrices. One way to avoid
this bound is to have $m_{\nu_e}\gg m_{\nu_{\mu}}$ so that the MSW
resonance condition is not met . However this possibility has the
apparent drawback that in this situation the MSW resonance condition
occurs for $\overline{\nu}_{\mu}\rightarrow \overline{\nu}_e$ leading to
contradictions with the celebrated $\overline{\nu}_e$ data of IMB and
Kamiokande experiments from SN1987A.
In  a recent paper[10], Fuller, Primack and Qian
have shown that this effect is not as strong as one might have
suspected, leaving this way to avoid the SN r-process bound as a viable
mechanism.

An arrangement of neutrino masses which satisfies the three inputs
above, as well as providing the two-neutrino version of the cold
plus hot dark matter, would have $m_{\nu_{\mu}}\ll m_{\nu_e}\simeq
m_{\nu_{\tau}}\approx 2.4$eV, with $\Delta m^2_{e\tau}\simeq 10^{-5}~eV^2$
for the large-angle MSW resolution to the solar neutrino deficit.
The r-process constraint is avoided by the inverted mass hierarchy.
If the $\nu_e$ was a Dirac particle, either of the two MSW solutions
could be used, but in theoretical frameworks driven by elegant symmetry
considerations, one is forced to use the large-angle solution, which
then leads to an automatic cancellation between the $\nu_e$ and $\nu_{\tau}$
contributions in the neutrinoless double beta decay amplitude in
equation (1). In this Letter,
we show that the relevant symmetry is a global $L_e-L_{\tau}$ symmetry
implemented in the framework of the left-right symmetric model, which
not only leads to near degeneracy between the $\nu_e$ and $\nu_{\tau}$
masses, but also it automatically satisfies the neutrinoless double beta
decay constraint, regardless of the  absolute values of those masses.
The scale of left-right symmetry in the simplest version of the model
has to be in the $10^{12}$ GeV range. In the low energy limit, this model
coincides with the standard electroweak theory
with the additional feature that
the neutrinos have the desired mass pattern. We later
point out some  phenomenological implications of the model.

\noindent{\bf 3. $L_e-L_{\tau}$ symmetry and $\nu_e$-$\nu_{\tau}$
 degeneracy implemented in a gauge model}

To see the main points of this discussion, let us ignore the muon neutrino
temporarily and consider only the $\nu_{e}$ and the $\nu_{\tau}$ with
the following mass matrix:

$$M_{e\tau}=\pmatrix{ \delta_1 & m\cr
                      m & \delta_2}.\eqno(2)$$

\noindent If we assume that $\delta_{1,2}\ll m$, then $L_e-L_{\tau}$
becomes a good symmetry of the model and the eigenvalues of this matrix
become nearly degenerate. Furthermore, the effective mass observed
in neutrinoless double beta decay in this case becomes
$$\langle m_{\nu_e}\rangle\simeq \delta ,$$
where $\delta$ is equal to $(\delta_1+\delta_2)/2$. Thus the double beta decay
constraint is easily satisfied. We can extend this discussion to the
case of three neutrinos easily as long as $m_{\nu_{\mu}}$ is very small.
The full $3\times 3$  matrix in this case would be
$$M=\pmatrix{-m\beta-\delta & -\mu_1 & m+\delta\cr
            -\mu_1 & \mu & -\mu_1 \cr
            m+\delta & -\mu_1 & m\beta-\delta}.\eqno(3)$$
 In the above matrix, we choose $m=2.4~eV$ to fit the HDM; $\delta\sim
10^{-5}~eV$ to fit the large-angle MSW solution;
$\beta\simeq (1-sin^22\theta_{e\tau})$ where $\theta_{e\tau}$ is the angle
required by the large-angle MSW solution to the solar neutrino problem. The
ratio $\mu_1/ m$ denotes the \num\-\nue\ mixing angle that can be measured
in the \num\ to \nue\ oscillation experiments; finally, the parameter
$\mu$ ( which is $\ll m$) is the \num\ mass.
Let us now proceed to the derivation of the full $3\times 3$ mass matrix
in a left-right symmetric gauge model.

 We will work within the conventional $SU(2)_L\times SU(2)_R\times U(1)_{B-L}$
models with the see-saw mechanism supplemented by a physical global
symmetry $U(1)_{L_e-L_{\tau}}$ which we will assume to be softly broken.
Let us start by
displaying only the leptonic and the relevant Higgs sectors. We denote
the lepton doublet by $\psi^T_i\equiv (\nu_i,e_i)$ ( where $i$ denotes
the generation index ) and assign the $\psi_L$
and $\psi_R$ to the left- and right-handed doublets under the gauge group.
For the Higgs sector, we choose the conventional multiplets of the usual
left-right symmetric model[11]; i.e., $\Delta_L(3,1,+2)\oplus \Delta_R
(1,3,+2)$ with $L_e-L_{\tau}$ quantum number zero; we choose two sets
of bi-doublets denoted by $\phi_{(0,1)}(2,2,0)$
with $L_e-L_{\tau}$ quantum numbers 0 and 1. As in Ref. 11,
the VEV of $\Delta^0_R$ breaks the $SU(2)_R$ symmetry.
We will assume this scale to be in the range of
$10^{11}$ to $10^{12}$ GeV.
The VEV's of $\phi_a$ break the standard model gauge symmetry. The
important point for us is that the potential minimization[11]
leads to an induced VEV for $\Delta^0_L$ given by $v_L\simeq \lambda
{{\langle\phi\rangle^2}\over{v_R}}$ which is now of order of a few eV's
for $\lambda\approx 10^{-2}$ to $10^{-1}$ ,
 where $v_L$ and $v_R$ are the VEV's of
$\Delta^0_L$ and $\Delta^0_R$, respectively, and $\lambda$ is a scalar
self coupling in the Higgs potential.

 To see the detailed structure of the neutrino masses, let us write
down the Yukawa couplings of the leptons and the Higgs multiplets:
$$L_Y=\Sigma_{i=1,2,3}~h_{ii}\overline{\psi}_{iL}\phi_0\psi_{iR}+
h_{12}\overline{\psi}_{eL}\phi_1\psi_{\mu R}+h_{23}\overline{\psi}_{\mu L}
\phi_1\psi_{\tau R}+$$ $$~~~~ [L\rightarrow R~ and~
 \phi_0\rightarrow \phi^{\dagger}_0~
and~ \phi_1\rightarrow \tau_2{\phi}^{T}_1\tau_2]$$
$$~~~~+[f_1 \psi_{eL}\psi_{\tau L}\Delta_L +
 f_2 \psi_{\mu L}\psi_{\mu L}\Delta_L
+ L\rightarrow R] +~ h.c. \eqno(4)$$
It is now easy to see that after symmetry breaking , one obtains
the following $6\times 6$ mass matrix in the basis  $[(\nu_e,\nu_{\mu},
\nu_{\tau}),( N_e, N_\mu, N_{\tau})]$  ( where we have denoted the
right-handed neutrinos by $N_i$):
$$M_B=\pmatrix{m_{LL} & m_{LR} \cr
               m^T_{LR} & M_{RR} }, \eqno(5a)$$
where
$$m_{LL}=\pmatrix{0 & 0 & f_1 v_L\cr
                  0 & f_2 v_L & 0 \cr
                  f_1v_L & 0 & 0 }; \eqno(5b)$$
$$m_{LR}=\pmatrix{h_{11}\kappa_0 & h_{12}\kappa_1 & 0\cr
                  h_{12}\kappa^{\prime}_1 & h_{22}\kappa_0 &h_{23}\kappa_1 \cr
                  0 & h_{23}\kappa^{\prime}_1 & h_{33}\kappa_0}; \eqno(5c)$$
and
$$M_{RR}=\pmatrix{0 & 0 & f_1v_R \cr
                  0 & f_2v_R & 0 \cr
                  f_1v_R & 0 & 0 }. \eqno(5d)$$
Here, we have denoted
$$\langle \phi_i \rangle= \pmatrix{\kappa_i & 0\cr
                                    0   & \kappa^{\prime}_i}. $$
Note that in deriving the above $m_{LR}$, we used the fact that
 under left-right transformation
we must have $\phi_1\rightarrow \tau_2{\phi_1}^{T}\tau_2$,
since $\phi_1$ possesses a non-zero $L_e-L_{\tau}$ quantum number.

 We see that the see-saw mechanism is
fully operative for all three neutrino generations, and the dominant mass
for the light neutrinos arises from the $v_L$
contributions . Since it conserves $L_e-L_{\tau}$ symmetry,
the mixing between the $\nu_e$ and $\nu_{\tau}$ is maximal. The $\nu_\mu$
also gets a mass from the $v_L$ term
 at this level. We choose $f_2\ll f_1$ so that
$m_{\nu_\mu}\ll eV$. The $\nu_e-\nu_{\tau}$ mass degeneracy is split
by the $L_e - L_{\tau}$ violating contributions that come from the Dirac
mass sector after the $\phi_{1,2}$ acquire VEV. An interesting point
is that if we choose the right-handed mass scale to be of order
$10^{12}$ GeV, then the \nue\-\nut\ mass splitting is naturally of
order $\approx {{m_e m_{\tau}}\over{v_R}}$ which is $\approx 10^{-5}$ eV
( for $v_R\simeq 10^{11}$ GeV ) as required by the large-angle MSW solution.
These $L_e-L_{\tau}$ violating terms also lead to the parameter
$\beta$ in Eq.(3), which can make the $sin^22\theta_{e\tau}$ slightly
less than one . The
$\nu_e-\nu_{\mu}$ mixing angle, however, comes naturally of the right
order of magnitude if we choose the parameters $h_{ij}$ suitably and
$\kappa_1\ll \kappa^{\prime}_1$.

\noindent{\bf 4. Phenomenology}

In this gauge model neutrinoless double beta decay is completely
unobservable, but since the $\nu_e$ mass is about half the present limit
from measurements at the endpoint energy of $^3H$ beta decay[12],
perhaps future work could test this hypothesis. The required large-angle
MSW resolution of the solar neutrino puzzle could also be tested by the
day-night effect in the solar neutrino flux when Super Kamiokande becomes
operative.

Since the $\nu_{\mu}-{\nu_{\tau}}$ mass splitting is large,
( i.e., $\Delta m^2_{\mu\tau} \simeq 6~eV^2$) it is
possible that the currently operating CHORUS and NOMAD experiments at CERN
or the future E803 experiment at Fermilab would detect neutrino oscillations.
It is especially intriguing that the preliminary LSND observation of
$\overline{\nu}_{\mu}\rightarrow\overline{\nu}_e$[13], for which also
$\Delta m^2_{e\mu}\simeq 6~eV^2$, may be compatible with this model.
Indeed, it was the primary motivation of two recent preprints[10,14]
in proposing this neutrino mass scheme to reconcile the LSND result
result with the r-process constraint. We wish to emphasize that the
LSND data could provide corroboration, but the mass matrix model is
motivated without this information.

The Raffelt-Silk preprint[14] also suggested a variation of our
scenario
 of three nearly degenerate neutrinos ($m_{\nu_e}\approx m_{\nu_{\mu}}
\approx m_{\nu_{\tau}}\approx 1.6~eV$)[15,16] but with the \nue\ and
\num\ masses inverted.
While this avoids the r-process bound and also provides for a \num\
$\rightarrow$\nut\ explanation of the atmospheric
 neutrino deficit, it does not satisfy the
constraint from neutrinoless double beta decay, unless one chooses a maximal
mixing angle scenario, as advocated in ref.17, with long-wavelength vacuum
oscillation to resolve the solar neutrino puzzle. This is also unlikely to
be borne out by LSND. The other preprint[14] also introduced a version
of the four-neutrino scheme (suggested by us earlier[15])
to provide an explanation of the atmospheric neutrino problem but again
inverting the \nue\ and \num\ masses, which makes the \nue\  a Dirac
neutrino. To accommodate a hot dark matter component, the atmospheric neutrino
deficit and a solar \nue\ deficiency, it is necessary to introduce a sterile
neutrino, if the scheme with three nearly degenerate neutrinos does not work.
The four-neutrino version we presented before[15] and which has
theoretically favored Majorana masses, might still avoid the supernova
constraints if the sterile neutrino alters the supernova dynamics sufficiently.

\noindent{\bf 5. Conclusion}

In summary, we have discussed the theoretical and phenomenological
consequences of an inverted mass hierarchy scenario for the three known
neutrinos that can account for the constraints on neutrino masses
and mixings: the solar neutrino deficit, mixed dark matter picture
of the universe, r-process generation of heavy elements, and
the lower limits on the lifetime for neutrinoless double beta decay.
This scenario does not account for the atmospheric neutrino deficit,
 which would have to have some
alternative explanation. We show that the conventional left-right
symmetric models with a  softly  broken
global $L_e-L_{\tau}$ symmetry and a high scale for the right-handed
symmetry breaking very naturally generate such mass matrices.
The day-night effects in the solar neutrino flux, as well as neutrino
data from future supernovae[18], can test such models. Earlier evidence
could come from the neutrino oscillation experiments such as LSND, CHORUS
and NOMAD.

One of the authors ( D. O. C.) would like to thank G. Fuller,
S. T. Petcov, J. R. Primack and G. Raffelt for discussions.

\vskip .4in

\centerline{References}

\item{[1].} R. Davis, {\it Prog. Part. Nucl. Phys.} {\bf 32}, 13 (1994);
            K. S. Hirata et al., \prd 44, 2241, 1991;
            Y. Suzuki, {\it Nucl. Phys. } {\bf B (Proc. Suppl.) 35}, 407
            (1994);
            J. N. Abdurashitov et al., \pl B238, 234, 1994;
            P. Anselman et al., \pl B327, 377, 1994.
\item{[2].} J. N. Bahcall and M. H. Pinsonneault, \rmp 64, 885, 1992;
            S. Turck-Chi\'eze and I. Lopes, {\it Ap. J.} {\bf 408}, 347 (1993).
\item{[3].} J. R. Primack, J. Holtzman, A. Klypin and D. O. Caldwell,
            \prl 74, 2160, 1995.
\item{[4].} K. S. Hirata et al., \pl B280, 146, 1992;
            Y. Fukuda et al.,\pl B335, 237, 1994;
            R. Becker-Szendy et al., \prd 46, 3720, 1992;
            P. J. Litchfield, {\it Proc. Int. Europhys. Conf.on High
            Energy Phys., Marseille}, 557 (1993) ( Edition Fronti\'eres,
            ed. J. Carr and M. Perottet).
\item{[5].} O. G. Ryazhskaya, Gran Sasso preprint LNGS-94/110
            (1994, unpublished)
\item{[6].} J. N. Bahcall, \pl B338, 276, 1994;
            P. I. Krastev and A. Yu. Smirnov, \pl B338, 282, 1994;
            V. Berezinsky, G. Fiorentini,and M. Lissia,{\it Phys. Rev.}
            {\bf D} ( in press) (1995);
            N. Hata and P. Langacker, {\it Phys. Rev.} {\bf D}
            (in press)(1995).
\item{[7].} For example, N. Hata and P. Langacker, \prd 50, 632, 1994;
            P. I. Krastev and S. T. Petcov, \prl 72, 1960, 1994.
\item{[8].} A. Balysh et al., {\it XXVIIth Int. Conf. on High Energy Phys.,
            Glasgow} ( 1994, to be published).
\item{[9].} Y-Z. Qian, G. M. Fuller and S. Woosley, \prl 71, 1965, 1993;
            Y-Z. Qian and G. M. Fuller, {\it Phys. Rev. D} (in press);
            G. Sigl, {\it Phys. Rev. D} (in press) (1995).
\item{[10].} G. M. Fuller, J. R. Primack and Y-Z. Qian, Preprint (1995)
             hep-ph 9502081.
\item{[11].} R. N. Mohapatra and G. Senjanovi\'c, \prd 23, 165, 1981.
\item{[12].} A. I. Belesev et al., {\it Inst. for Nucl. Research preprint
            INR 862/94} (unpublished).
\item{[13].} W. C. Louis, {\it Nucl. Phys.}{\bf B( Proc. Suppl.) 38}, 229
             (1995); D. O. Caldwell, {\it ibid.}, 394
\item{[14].} G. Raffelt and J. Silk, Berkeley preprint, (1995).
\item{[15].} D. O. Caldwell and R. N. Mohapatra, \prd 48, 3259, 1993
\item{[16].} D. O. Caldwell and R. N. Mohapatra  \prd 50, 3477, 1994.
\item{[17].} R. N. Mohapatra and S. Nussinov, {\it Phys. Lett. B}
             (to appear) (1995).
\item{[18].} A. Yu. Smirnov, D. N. Spergel and J. N.
              Bahcall, \prd 49, 1389, 1994.

\end

 The Higgs sector of the model
consists of the triplet fields $\Delta_L(3,1,+2)\oplus\Delta_R(1,3,+2)$
and the bidoublet field $\phi(2,2,0)$. The only point worth mentioning
about the rest of the model is the fact the $\nu_{\mu}$ mass of $2.4$ eV
will be generated by the induced vacuum expectation value(VEV) for
$\Delta^0_L\equiv v_L$ which arises from potential minimization\refto{16}
of the model and is given by $v_L\simeq \lambda {{\kappa^2}\over{v_R}}$
($\kappa$ is the VEV of the $\phi$ field and corresponds to the scale of
the standard electroweak model. Note that all see-saw induced neutrino masses
are in the range of $10^{-2}$ eV or so. We will assume that all other
$v_L$ induced neutrino masses are in the milli-eV range.
This kind of scheme is more easily embedded into GUT theories such as
$SO(10)$, which would then explain the origin of the high scales
in the neutrino see-saw formula. In this note note, we will content
ourselves only with the left-right symmetric model.

\end